\documentclass[twocolumn,secnumarabic,amssymb, nobibnotes, aps,superscriptaddress]{revtex4-1}
\usepackage{amsmath}
\usepackage{graphicx}
\usepackage{bm}

\begin{document}
\title{Tunable magneto-optical properties of single-layer tin diselenide: From GW approximation to large-scale tight-binding calculations}%
\author{Hongxia Zhong}
\affiliation{Key Laboratory of Artificial Micro- and Nano-structures of Ministry of Education and School of Physics and Technology, Wuhan University, Wuhan 430072, China}
\author{Jin Yu}%
\affiliation{Institute for Molecules and Materials, Radboud University, Heyendaalseweg 135, 6525AJ Nijmegen, Netherlands}
\author{Kaixiang Huang}%
\affiliation{Key Laboratory of Artificial Micro- and Nano-structures of Ministry of Education and School of Physics and Technology, Wuhan University, Wuhan 430072, China}
\author{Shengjun Yuan}%
\email{s.yuan@whu.edu.cn}
\affiliation{Key Laboratory of Artificial Micro- and Nano-structures of Ministry of Education and School of Physics and Technology, Wuhan University, Wuhan 430072, China}
\date{\today}
\begin{abstract}
	A parameterized tight-binding (TB) model based on the first-principles GW calculations is developed for single layer tin diselenide (SnSe$_2$) and used to study its electronic and optical properties under external magnetic field. The truncated model is derived from six maximally localized wannier orbitals on Se site, which accurately describes the quasi-particle electronic states of single layer SnSe$_2$ in a wide energy range. The quasi-particle electronic states are dominated by the hoppings between nearest wannier orbitals ($t_1$-$t_6$). Our numerical calculation shows that, due to the electron-hole asymmetry, two sets of Landau Level spectrum are obtained when a perpendicular magnetic field is applied. The Landau Level spectrum follows linear dependence on the level index and magnetic field, exhibiting properties of two-dimensional electron gas in traditional semiconductors. The optical conductivity calculation shows that the optical gap is very close to the GW value, and can be tuned by external magnetic field. Our proposed TB model can be used for further exploring the electronic, optical, and transport properties of SnSe$_2$, especially in the presence of external magnetic fields.
\end{abstract}
\maketitle

\section{Introduction}
Two-dimensional (2D) transition metal dichalcogenides (TMDCs) have been receiving continuous concerning in the last few years because of their potential applications in electronic and optical devices\cite{wang2012electronics,jariwala2014emerging,zhong2016interfacial}. More interestingly, their metastable 1T phases are reported to have some exotic properties\cite{cercellier2007evidence,rohaizad20171t}. Thus, the main group metal dichalcogenide (MDC) family, whose ground state holds the same atomic configuration of 1T TMDCs, has attracted much attention recently\cite{zhang2015few,zhou2016large,zhou2018tunneling,ying2018unusual}. These layered MDCs are highly abundant in earth, environment friendly, and low toxicity\cite{su2014chemical,burton2016electronic,hu2017innovative}. Particularly, the 1T structure endows them with unique anisotropic thermal transport properties\cite{xu2017anisotropic,chen2018enhanced}. One typical example is tin diselenide (SnSe$_2$), whose ZT value reaches 2.95 (0.68) along a (c) axis, much higher than that in 2H- and 1T-TMDCs\cite{luo2018n,li2016thermoelectric}. In addition to the high thermoelectric performance, they also exhibit excellent electronic properties. The field effect transistor made of few layer SnSe$_2$ is reported to have high current on/off ratio (10$^5$) and mobility (85 cm$^2$ V$^{-1}$ S$^{-1}$)\cite{guo2016field,pei2016few}, respectively. When it is combined with other 2D materials like WSe$_2$ and black phosphorus to form van der Waals (vdW) heterostructures\cite{na2019gate,roy20162d,yan2017tunable}, negative differential resistance and good subthreshold swing phenomenon are observed. The large work function and broken band gap alignment with WSe$_2$ make it an ideal candidate for high efficiency two-dimensional heterojunction interlayer tunneling field effect transistors (TFETs)\cite{roy20162d}. Very recently, gate-induced superconductivity is demonstrated in 1T SnSe$_2$ through the ionic liquid gating technique\cite{zeng2018gate}. Most of these exciting results are concluded from high-quality layered SnSe$_2$, which now can be obtained by chemical vapor deposition (CVD) \cite{zhou2015ultrathin} and physical mechanical exfoliation from the bulk crystals\cite{su2013snse2} in experiments. However, how to better describe the physical phenomena of realistic SnSe$_2$ system comparable to experimental samples is still under exploring.

Theoretically, the physical properties of single layer SnSe$_2$ have been investigated by first-principles calculations\cite{gonzalez2016layer,li2016thermoelectric,shafique2017ultra}. One disadvantage of such method is the high computational cost, which can only handle systems with nonequivalent atomic system less than 10$^3$. Thus, it is not enough to describe the disordered, inhomogeneous materials, and twisted multilayer materials at large scales. Alternately, the method of model Hamiltonians can address this problem with large systems up to 10$^9$ atoms. And the tight-binding (TB) method is much more efficient and flexible but less transferable. Among 2D materials, several TB models are available to capture the relevant physical properties in graphene\cite{reich2002tight} and its derivatives\cite{mazurenko2016role}, black phosphorous\cite{rudenko2014quasiparticle,rudenko2015toward}, monolayer antimony\cite{rudenko2017electronic}, arsenene\cite{yu2018tunable}, and TMDCs\cite{liu2013three,cappelluti2013tight}. A very recent work has shown success in investigating the tunable magneto-optical properties of monolayer SnS$_2$ from maximally localized wannier function. As a member of the same group MDCs, we would expect more interesting physical properties of single layer SnSe$_2$\cite{yu2018effective}. Moreover, many-body effects are crucial for 2D SnSe$_2$ due to the depressed screening and reduced dimensionality of suspended 2D semiconductors. It is thus helpful to develop an effective Hamiltonian for single layer SnSe$_2$ based on the quasi-particle energy as well, to further study its physical properties under external field.

In this paper, we derive a tractable TB model for single layer SnSe$_2$ based on six orbitals with trigonal rotational symmetry. The model describes the quasi-particle electronic states in a wide spectrum region, which are validated by good consistence with the band structure and density of states from first-principles calculations. In the presence of a perpendicular magnetic field, two sets of Landau levels appear, which disperse linearly with the level index $n$ and magnetic field strength $B$. When the value of $B$ increases, the optical conductivity spectrum of single layer SnSe$_2$ shows a blue shift, accompanied by a broadening of the gap. Our proposed TB model paves a new way in considering real SnSe$_2$ system with disorders, many body effects, and applied fields. 

The remainder of this paper is organized as follows: In Sec. II, we introduce the calculation method and converge parameters. In Sec. III, the atomic structure and quasi-particle electronic properties of single layer SnSe$_2$ are presented. In Sec. IV, we propose our simple TB model and analyze the fitting results. In Sec. V, we discuss the effect of magnetic field on single layer SnSe$_2$ based on the developed TB model. Finally, we give a summary of our current work in Sec. VI.

\section{Computational Methods}
In order to get a reliable TB model, the geometrical optimization and electronic properties of monolayer SnSe$_2$ are performed by first-principles calculations. We fully relax the atomic structures according to the force and stress performed by density functional theory (DFT) using the Vienna ab initio calculation package (VASP) code\cite{kresse1996efficiency,kresse1999ultrasoft}. The generalized gradient approximation (GGA) functional of the PBE form\cite{perdew1996generalized} and the projected augmented-wave method (PAW)\cite{blochl1994projector} are adopted. The cutoff energy is set to 500 eV after convergence tests. An equivalent Monkhorst-Pack $k$-points\cite{monkhorst1976special,pack1977special} grid 15 $\times$ 15 $\times$ 1 is chosen for relaxation and 40 $\times$ 40 $\times$ 1 for the static calculations. In our current calculations, the total energy is converged to less than 10$^{-5}$ eV, and the maximum force is less than 0.02 eV/$\mathrm{\AA}$ during the optimization. A vacuum layer of 30 $\mathrm{\AA}$ is fixed to avoid spurious interactions. 

The quasi-particle energies and band gaps are calculated by the GW approximation within the general plasmon pole model\cite{rohlfing2000electron}. The involved unoccupied conduction band number for calculating the dielectric function and self-energy is about ten times the occupied valence band number. All the GW calculations are performed with the BERKELEYGW code\cite{deslippe2012berkeleygw} including the slab Coulomb truncation scheme to mimic suspended monolayer structures\cite{ismail2006truncation,rozzi2006exact}.

The construction of the Wannier functions and TB parametrization of the DFT Hamiltonian are done with the WANNIER90 code\cite{mostofi2008wannier90}. The obtained hopping parameters are further discarded and re-optimized through a least-squares fitting of the band structure. The electronic density of states and the optical conductivity with external magnetic field are calculated by the tight-binding propagation method (tipsi) \cite{yuan2010modeling,yuan2011landau,yuan2012enhanced}. This numerical method is based on the Chebyshev polynomial algorithm without the diagonalization of the Hamiltonian matrix. It is thus an efficient numerical tool in large-scale calculations of quantum systems with more than millions of atoms.

\section{first-principles resutls}
\begin{figure*}[pht]	
	\includegraphics[width=13.0cm]{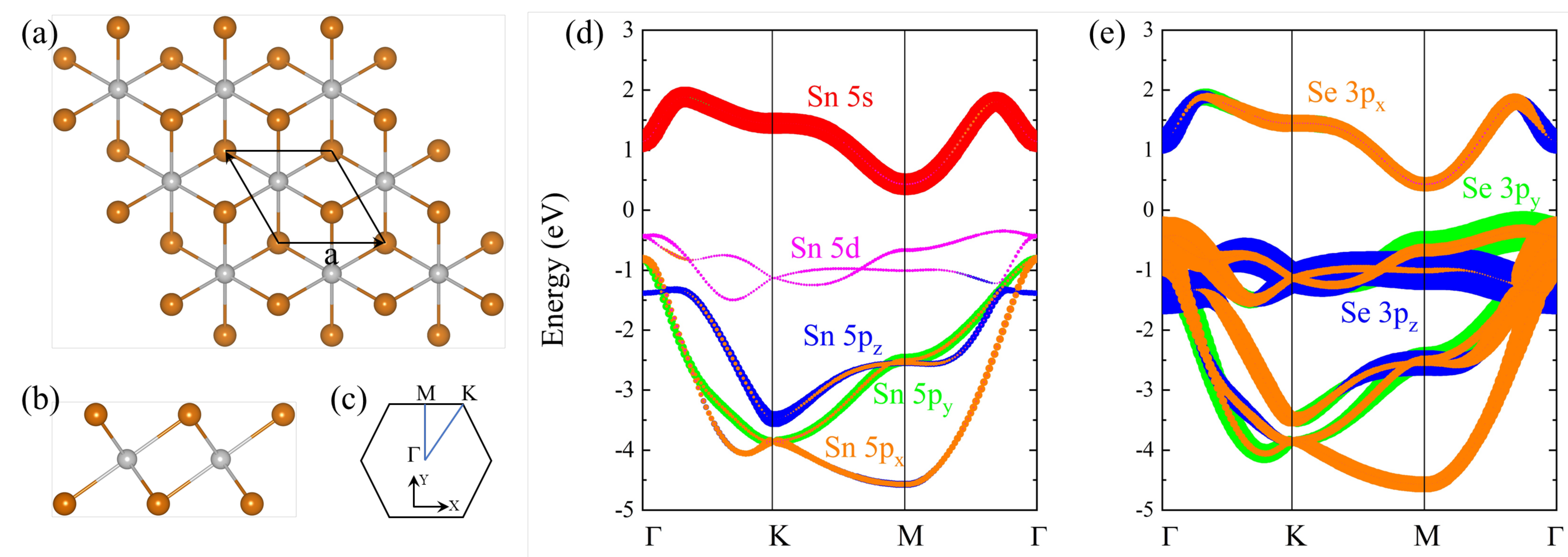}
	\caption{(a)-(b) Top and side views of atomic structure of single layer SnSe$_2$. The primitive cell and lattice vector $\textbf{a}$ are indicated. (c) The first Brillouin zone of single layer SnSe$_2$. (d)-(e) Orbital decomposed band structure of Sn and Se in single layer SnSe$_2$, with the corresponding $s$, $p_x$, $p_y$, $p_z$, and $d$ orbitals being red, orange, green, blue, and magenta, respectively. The size of the symbol represents the orbital weight.}
	\label{fig:label1}
\end{figure*}
Unlike traditional TMDCs with $D_{3h}$ space group, single layer SnSe$_2$ crystallizes in a 1T-phase structure with $D_{3d}$ point group, as presented in Figs.~\ref{fig:label1}(a)-(b). It includes a threefold rotation symmetry axis $C_{3v}$ and a vertical mirror plane $\delta_d$. The optimized lattice parameter \emph{\textbf{a}} is 3.869 {\AA}, the bond length of Sn-Se is 2.751 {\AA}, and the distance between two Se plane is 3.193 {\AA}. They are in line with previous results\cite{huang2016first,gonzalez2016layer}. The projected band structures of Sn and Se atoms in single layer SnSe$_2$ are presented in Figs.~\ref{fig:label1}(d)-(e). Single layer SnSe$_2$ is an indirect-gap semiconductor; the conduction band minimum (CBM) is located at the M point, while the valence band maximum (VBM) is slightly away from the $\Gamma$ point (energy difference $\sim$85 meV). The PBE-calculated indirect band gap between the $\Gamma$ and M points is about 1.580 eV, consistent with a reported value of 1.690 eV\cite{gonzalez2016layer}. Our result shows that the presented six bands in low energy window mainly involve the $s$ and $p$ orbitals of Sn and Se atoms. The first and second valence bands at the $\Gamma$ point are doubly degenerate, with orbital decomposition of the corresponding wave function $\left | \psi^{VBM}(\Gamma) \right \rangle$ = 0.307 $\left | p_x \right \rangle$$_\mathrm{{Se}}$+0.952 $\left | p_y \right \rangle$$_\mathrm{{Se}}$. And the wave function of the CBM is  $\left | \psi^{CBM}(\mathrm{M}) \right \rangle$ = 0.789 $\left | s \right \rangle$$_\mathrm{{Sn}}$+ 0.535 $\left | p_x \right \rangle$$_\mathrm{{Se}}$ +  0.179 $\left | p_y \right \rangle$$_\mathrm{{Se}}$ +0.241
$\left | p_z \right \rangle$$_\mathrm{{Se}}$.

It is well known that DFT usually underestimates the band gap of semiconductors. We therefore have performed the GW calculation to get reliable quasi-particle band gap of monolayer SnSe$_2$. Similar to those found in other monolayer 2D semiconductors, such as monolayer hexagonal TMDCs\cite{zhong2015quasiparticle} and phosphorene\cite{tran2014layer}, significant self-energy enhancements are observed in SnSe$_2$; at the “single-shot” G$_0$W$_0$ level, the quasi-particle band gap at $\Gamma$ point of monolayer SnSe$_2$ is increased to be 2.750 eV. Self-consistent GW (sc-GW) scheme beyond single-shot calculations has been verified to be necessary for some 2D semiconductors. Hence, we perform one self-consistent update to the Green’s function G; the quasiparticle band gap is further increased to be 3.069 eV. More self-consistent steps only slightly change the band gap ( $\sim$0.1eV) and we stop at the sc-G$_1$W$_0$ level. Finally, the enhanced many-electron effects enlarge the band gaps, but do not change the dispersion shape. The energy difference between the VBM and $\Gamma$ point is about $\sim$77 meV, which is nearly the same with the PBE-calculated value. In the following, all our discussions are based on the finalized sc-G$_1$W$_0$ results.

\section{Tight-binding model}
\begin{figure}[ht]	
	\includegraphics[width=8.5cm]{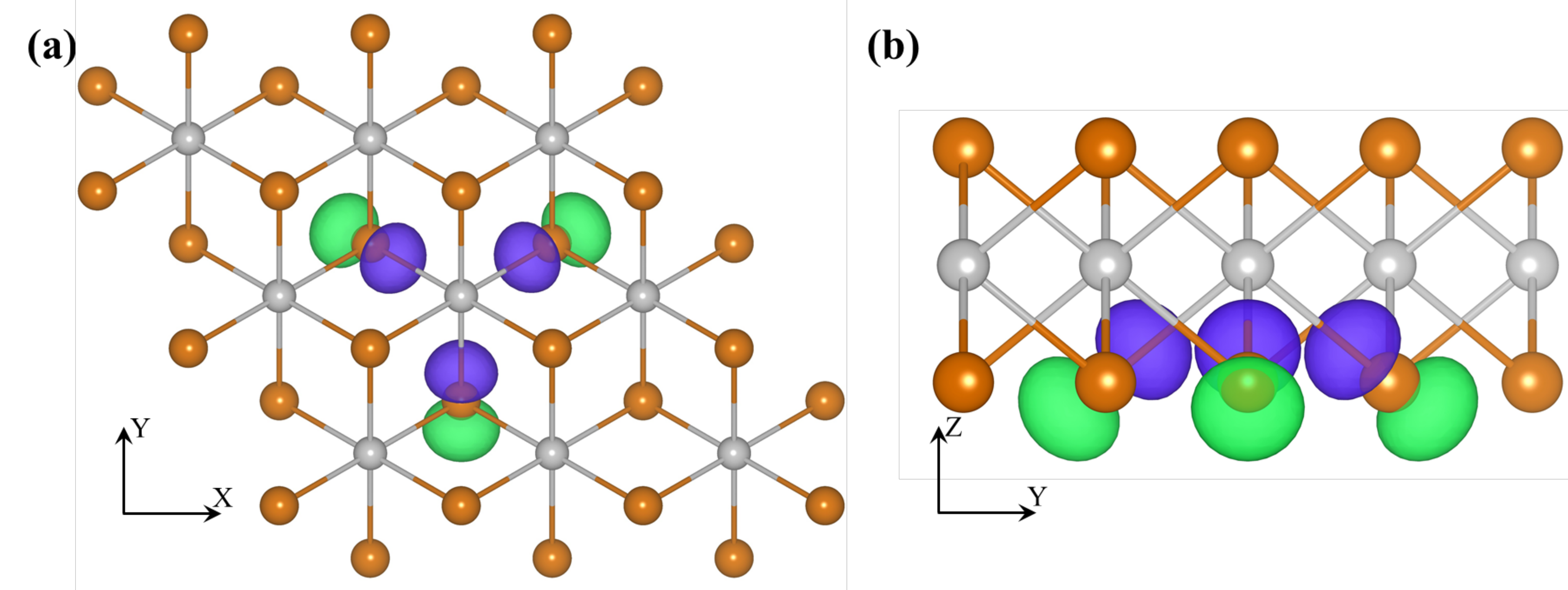}
	\caption{Maximally Localized Wannier Functions of single layer SnSe$_2$ corresponding to the basis of the TB Hamiltonian presented in this work. For clarity, orbitals are shown for one sublattice with one orbital per atom only. The orbitals in the second sublattice are symmetric with respect to the inversion center.}
	\label{fig:label2}
\end{figure}

Given that the valence and conduction bands are dominated by $s$ and $p$ orbitals, and that they are separated from other states, it is possible to provide an accurate description of those states in terms of a tractable TB model in the low-energy region. Our parametrization procedure used in this work is based on the formalism of Maximally Localized Wannier Functions (MLWFs), and six orbitals are obtained as the basis for the TB model. A real-space distribution of the MLWFs obtained for single layer SnSe$_2$ is shown in Fig.~\ref{fig:label2}, where a combination of three orbitals localized on each Se atom, giving rise to six MLWFs per cell on two sublattices. The three orbitals are equivalent with a rotational symmetry of 2$\pi$/3, effectively reducing the independent TB parameters.

The resulting non-relativistic TB model is given by a 6 by 6 effective Hamiltonian. The matrix elements of the Hamiltonian in Bloch-representation is 
\begin{equation}
\begin{aligned}
H_{ \alpha\beta} = \langle \alpha, \mathrm{k} |H| \beta, \mathrm{k}\rangle & = \sum_{\bm{\mathrm{R}}} e^{i\bm{\mathrm{k \cdot R}}} \langle \alpha, 0 |H| \beta, \bm{\mathrm{R}}\rangle \\
& = t_{ \alpha\beta}(\bm{\mathrm{R}})\sum_{\bm{\mathrm{R}}} e^{i\bm{\mathrm{k \cdot R}}}
\label{Hamiltonian matrix1}
\end{aligned}
\end{equation}

\begin{figure}[ht]	
	\includegraphics[width=8.5cm]{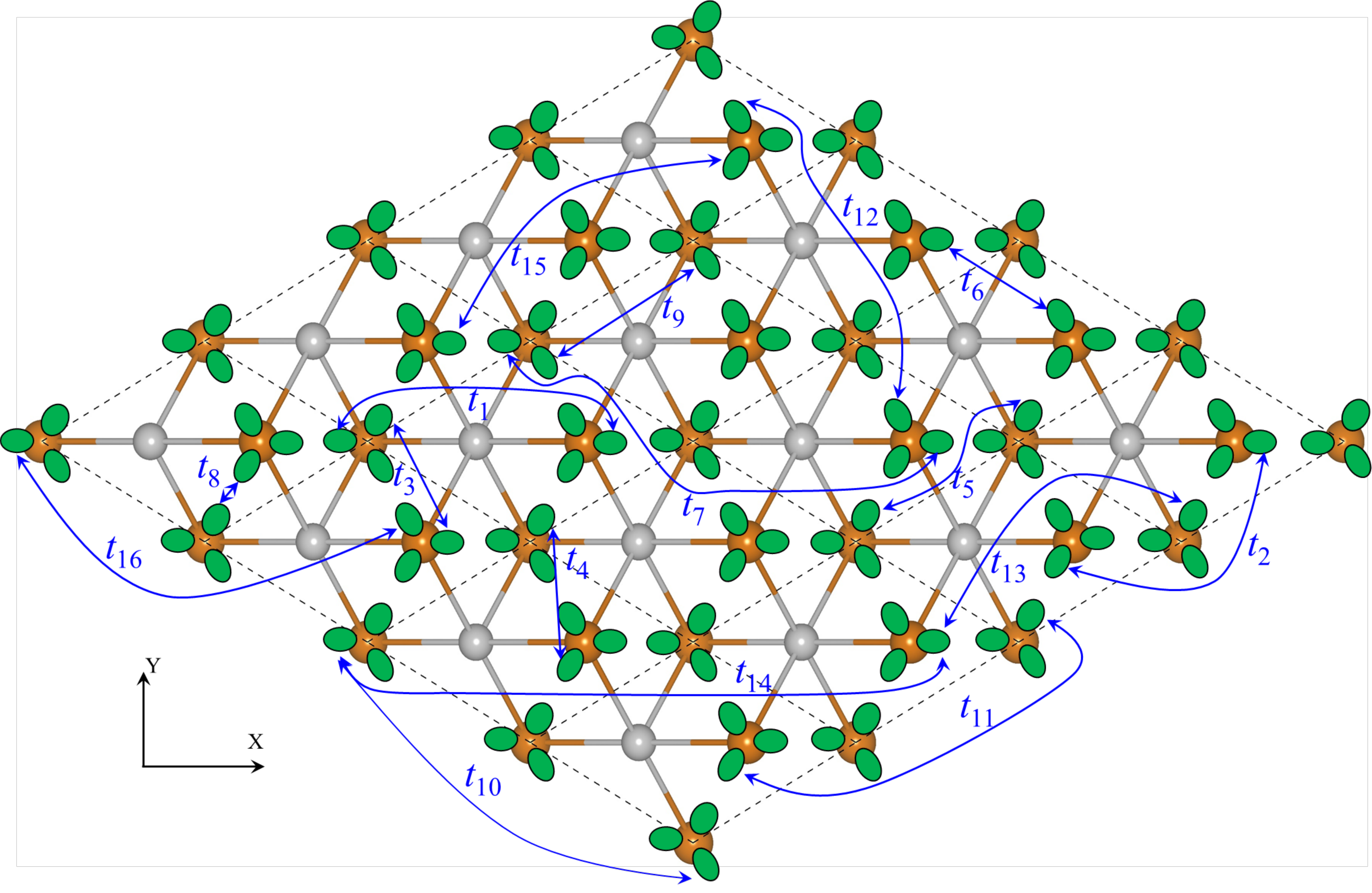}
	\caption{Hopping diagram in the TB model of single layer SnSe$_2$. The orbitals are represented by the negative part of the wannier orbitals as indicated by the green ellipses.}
	\label{fig:label3}
\end{figure}

Where t$_{\alpha\beta}$ is the effective hopping parameter describing the interaction between $\alpha$ and $\beta$ orbitals residing at central and neighbour atoms, respectively. To make the model more tractable yet accurate enough, we first discard the hoppings with an interatomic distance larger than 11.620 {\AA}, then ignore hopping parameters with amplitudes $|t|$ $<$ 19 meV. The residual hopping parameters are furthre re-optimized by minimizing the following least square function 
\begin{equation}
\delta(\{t_i\})=\sum_{n,\bm{\mathrm{k}}}\frac{[E_{n,\bm{\mathrm{k}}}^{G_1W_0}(\{t_i\})^2-E_{n,\bm{\mathrm{k}}}^{TB}(\{t_i\})^2]}{exp[(E_{n,\bm{\mathrm{k,V/CB}}}^{TB}(\{t_i\})-E_{\bm{\mathrm{V/CBM}}}^{G_1W_0}(\{t_i\}))]^2}
\end{equation}

where $\{t_i\}$ is the hopping in Eq.~(\ref{Hamiltonian matrix1}), $E_{n,\bm{\mathrm{k}}}^{G_1W_0}(\{t_i\})$  ($E_{n,\bm{\mathrm{k}}}^{TB}(\{t_i\})$) corresponds to the eigenvalues of G$_1$W$_0$ (TB) Hamiltonian with $n$ and $\bm{\mathrm{k}}$ being the band index and momenta along the high-symmetry in the first Brillouin zone, respectively. 
In order to get much more reliable results near the band edge, we add a Gaussian function
$exp[(E_{n,\bm{\mathrm{k,V/CB}}}^{TB}(\{t_i\})-E_{\bm{\mathrm{V/CBM}}}^{G_1W_0}(\{t_i\})]^2$ here, in which
$E_{\bm{\mathrm{V/CBM}}}^{G_1W_0}(\{t_i\})$ is the energy of VBM/CBM from G$_1$W$_0$ approximation. The remaining orbitals and relevant hopping parameters are schematically shown in Fig.~\ref{fig:label3} and Table \ref{table:label1}. 

\begin{table}
	\caption{Hopping parameters $t_i$ (in eV) assigned to the simple TB Hamiltonian of SL-SnSe$_2$. $d$ is the distance between the atomic sites on which the interacting orbitals are centered. The hoppings are schematically shown in Fig.~\ref{fig:label3}}
	\begin{ruledtabular}
		\begin{tabular*}{7cm}{cccccccccccccccc}
			&$i$     &$t_i (\mathrm{eV})$    &$d (\mathrm{\AA})$   &$i$      &$t_i (\mathrm{eV})$    &$d (\mathrm{\AA})$     &$i$      &$t_i (\mathrm{eV})$    &$d (\mathrm{\AA})$    \\
			\hline
			&1       &-1.513       &5.491        &7       &-0.093     &8.664            &13    &0.024    &6.717         \\
			&2       &-0.681       &3.869        &8       &-0.093     &3.897            &14    &-0.020    &11.617         \\
			&3       &0.636        &3.897        &9       &-0.058     &3.869            &15    &0.047     &7.738          \\
			&4       &0.399        &3.897        &10      &-0.041     &7.738            &16    &-0.019    &8.664           \\ 
			&5       &0.285        &3.869        &11      &0.076      &6.717           &      &         &           &   \\
			&6       &0.300        &3.869        &12      &0.056      &6.701          &      &         &           &     \\
		\end{tabular*}
	\end{ruledtabular}
    \label{table:label1}
\end{table}

Considering the inversion symmetry of atomic structure, our reciprocal space  Hamiltonian matrix in Eq.~(\ref{Hamiltonian matrix1}) can be represented as

\begin{equation}
H(\bm{\mathrm{k}}) = \left(
\begin{array}{cc}
E(\bm{\mathrm{k}})  &T(\bm{\mathrm{k}})  \\
T^\dagger(\bm{\mathrm{k}})  &E(\bm{\mathrm{k}}_r) 
\label{}
\end{array}
\right) 
\end{equation}
where  $E(\bm{\mathrm{k}})$ and $T(\bm{\mathrm{k}})$ are 3 $\times$ 3 matrices describing the intra- and inter-sublattice interactions, respectively. The subscript $r$ in U($k_r$) indicates rotation in the opposite direction, resulting from the vertical mirror symmetry $\delta_d$. Because the three basic orbitals have threefold rotation symmetry $C_{3v}$, the corresponding matrices have the form 

\begin{equation}
E(\bm{\mathrm{k}}) = \left(
\begin{array}{ccc}
A(\bm{\mathrm{k}})                   &B(\bm{\mathrm{k}})      &B^\star(\bm{\mathrm{\overline{\overline{k}}}}) \\
B^\star(\bm{\mathrm{{k}}})  &A(\bm{\mathrm{\overline{k}}})     &B(\bm{\mathrm{\overline{k}}}) \\
B(\bm{\mathrm{\overline{\overline{k}}}})   &B^\star(\bm{\mathrm{\overline{k}}})    &A(\bm{\mathrm{\overline{\overline{k}}}}) 
\label{Hamiltonian matrix2}
\end{array}
\right), 
\end{equation}

and 
\begin{equation}
T(\bm{\mathrm{k}}) = \left(
\begin{array}{ccc}
C(\bm{\mathrm{k}})                   &D(\bm{\mathrm{k}})                &C(\bm{\mathrm{\overline{\overline{k}}}}) \\
D(\bm{\mathrm{{\overline{k}}}})                 &C(\bm{\mathrm{k}})     &C(\bm{\mathrm{\overline{k}}}) \\
C(\bm{\mathrm{\overline{k}}})     &C(\bm{\mathrm{\overline{\overline{k}}}})   &D(\bm{\mathrm{\overline{\overline{k}}}})
\label{Hamiltonian matrix3}
\end{array}
\right). 
\end{equation}
			
where $\bm{\mathrm{\overline{k}}}$ and $\bm{\mathrm{\overline{\overline{k}}}}$ are the $\bm{\mathrm{k}}$ vector rotated by 2$\pi$/3 and 4$\pi$/3, respectively. The matrix elements in Eqs.~(\ref{Hamiltonian matrix2})and (\ref{Hamiltonian matrix3}) are

\begin{equation}
\begin{aligned}
A(\bm{\mathrm{k}})  = & 2t_5\cos(\frac{\sqrt{3}}{2}ak_\mathrm{x}-\frac{{1}}{2}ak_\mathrm{y})+ 2t_5\cos(\frac{\sqrt{3}}{2}ak_\mathrm{x}+\frac{{1}}{2}ak_\mathrm{y})\\
&+2t_{9}\cos(ak_\mathrm{y})+2t_{12}\cos(\sqrt{3}ak_\mathrm{x})
\label{Hamiltonian2}
\end{aligned}
\end{equation}

\begin{equation}
\begin{aligned}
B(\bm{\mathrm{k}}) = &t_2e^{i(\frac{\sqrt{3}}{2}ak_\mathrm{x}-\frac{{1}}{2}ak_\mathrm{y})}+t_6e^{i(-\frac{\sqrt{3}}{2}ak_\mathrm{x}+\frac{{1}}{2}ak_\mathrm{y})}\\
&+t_{10}e^{i(\sqrt{3}ak_\mathrm{x}-ak_\mathrm{y})}+t_{15}e^{i(-\sqrt{3}ak_\mathrm{x}+ak_\mathrm{y})}
\label{Hamiltonian3}
\end{aligned}
\end{equation}

\begin{equation}
\begin{aligned}
C(\bm{\mathrm{k}}) = &t_3e^{-iak_\mathrm{y}}+t_{13}e^{i(\frac{\sqrt{3}}{2}ak_\mathrm{x}+\frac{1}{2}ak_\mathrm{y})}+t_{13}e^{i(-\sqrt{3}ak_\mathrm{x}-ak_\mathrm{y})}\\
&+t_{16}e^{i(-\sqrt{3}ak_\mathrm{x}-2ak_\mathrm{y})}+t_{16}e^{i\sqrt{3}ak_\mathrm{x}}
\label{}
\end{aligned}
\end{equation}

\begin{equation}
\begin{aligned}
D(\bm{\mathrm{k}}) = &t_1e^{i(\frac{\sqrt{3}}{2}ak_\mathrm{x}-\frac{1}{2}ak_\mathrm{y})}+t_4(1+e^{-iak_\mathrm{y}}) \\
&+t_7e^{i(\sqrt{3}ak_\mathrm{x}-ak_\mathrm{y})}+t_7e^{i\sqrt{3}ak_\mathrm{x}} \\
&+t_8e^{i(-\frac{\sqrt{3}}{2}ak_\mathrm{x}-\frac{1}{2}ak_\mathrm{y})}+t_{11}e^{i(\frac{\sqrt{3}}{2}ak_\mathrm{x}+\frac{1}{2}ak_\mathrm{y})} \\
&+t_{11}e^{i(\frac{\sqrt{3}}{2}ak_\mathrm{x}-\frac{3}{2}ak_\mathrm{y})}+t_{14}e^{i(\frac{3\sqrt{3}}{2}ak_\mathrm{x}-\frac{1}{2}ak_\mathrm{y})}
\label{}
\end{aligned}
\end{equation}

\begin{figure}[ht]	
	\includegraphics[width=8.5cm]{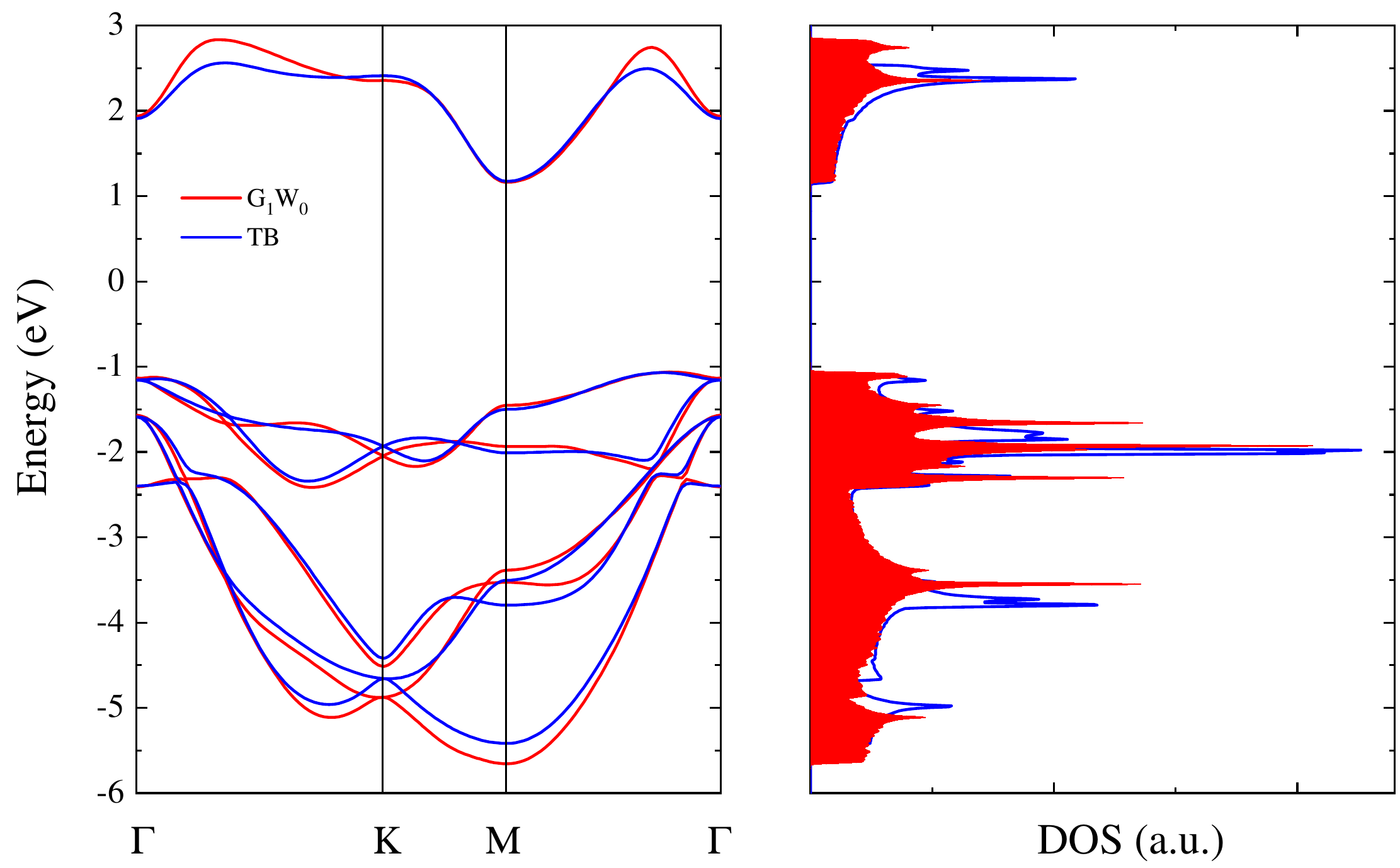}
	\caption{Comparison of the band structure and density of states for single-layer SnSe$_2$ obtained from G$_1$W$_0$ (red line) and TB (blue line) calculations, respectively.}
	\label{fig:label4}
\end{figure}

The band structure and density of states (DOS) obtained from the given TB model are shown in Fig.~\ref{fig:label4}. The quasi-particle energy within G$_1$W$_0$ is plotted for comparison. One can see a good match between the TB and original first-principles calculations, especially for the band edges. This can be further quantified by the band gaps and effective masses analysis, which are accurately reproduced by the given TB model as shown in Table \ref{table:label2}. The indirect (direct) band gaps obtained from TB model are 2.331 (3.066) eV, in good agreement with the value of 2.295 (3.070) eV in G$_1$W$_0$ calculation. This consistence is also remarkable in the effective masses of carriers in single-layer SnSe$_2$. The anisotropic effective masses of electrons along the K-M and $\Gamma$-M direction are 0.558 $m_0$ and 0.202 $m_0$ in the proposed TB model, and the corresponding values from G$_1$W$_0$ are 0.650 $m_0$ and 0.202 $m_0$, respectively. Thus, our proposed TB model can be used effectively in describing the electronic, optical and transport properties of single layer SnSe$_2$. 

\begin{table}
	\caption{Indirect ($\Gamma$M) and direct ($\Gamma$$\Gamma$) band gaps $E_g$ (in eV), and effective masses $m$ (in units of free electron mass $m_0$) for electrons and holes in single layer SnSe$_2$ at high-symmetry points in the first Brillouin zone calculated by the DFT+GW and TB model presented in this work.}
	\begin{ruledtabular}
		\begin{tabular*}{7cm}{ccccccc}
		\multicolumn{1}{c}{} &\multicolumn{2}{c}{Energy gap}  & \multicolumn{2}{c}{Electrons}  &\multicolumn{2}{c}{Holes} \\
	    \multicolumn{1}{c}{Method}       &\multicolumn{1}{c}{GM}          & \multicolumn{1}{c}{GG}  &\multicolumn{1}{c}{KM}   &\multicolumn{1}{c}{GM}      &\multicolumn{1}{c}{GK}  &\multicolumn{1}{c}{GM}  \\
	    \hline
		DFT+GW    &2.295        &3.070    &0.650     &0.202       &1.133  &0.913        \\
	    TB        &2.331        &3.066    &0.558     &0.202       &1.261  &0.854      \\
		\end{tabular*}
	\end{ruledtabular}
    \label{table:label2}
\end{table}

\section{magneto-electronic properties}
\begin{figure}[ht]	
	\includegraphics[width=8.0cm]{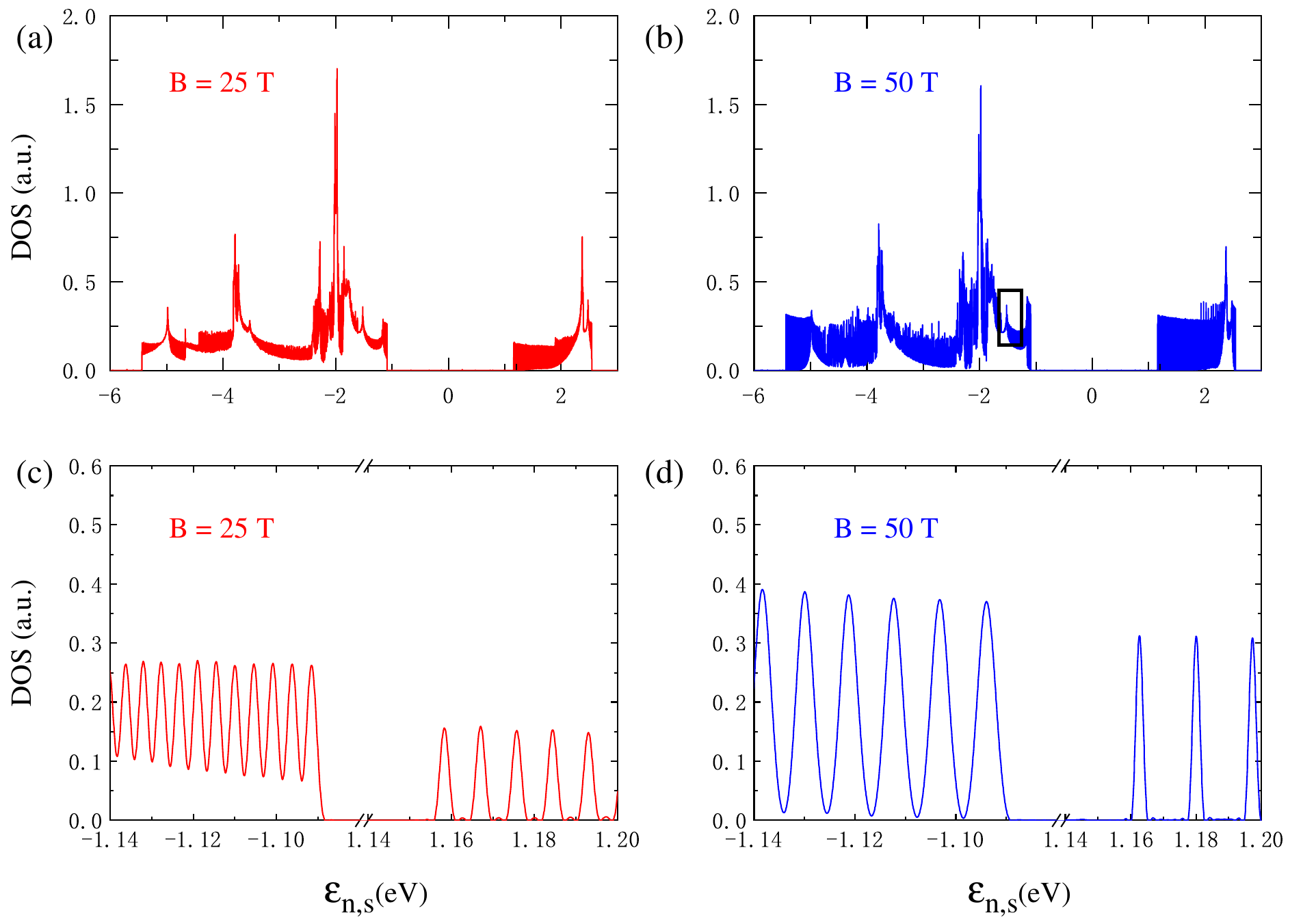}
	\caption{The Landau level spectrum of single layer SnSe$_2$ under uniform magnetic field of (a) 25T, and (b) 50T, respectively. We zoom in the band edge parts of (a) and (b), which are shown at (c) and (d), respectively. The sample used in the numerical calculations contains 2 $\times$ 1500 $\times$ 1500 atomic sites, and we use the periodic boundary conditions (XY) in the plane of SnSe$_2$ layers.}
	\label{fig:label5}
\end{figure}

\begin{figure}[ht]	
	\includegraphics[width=7.5cm]{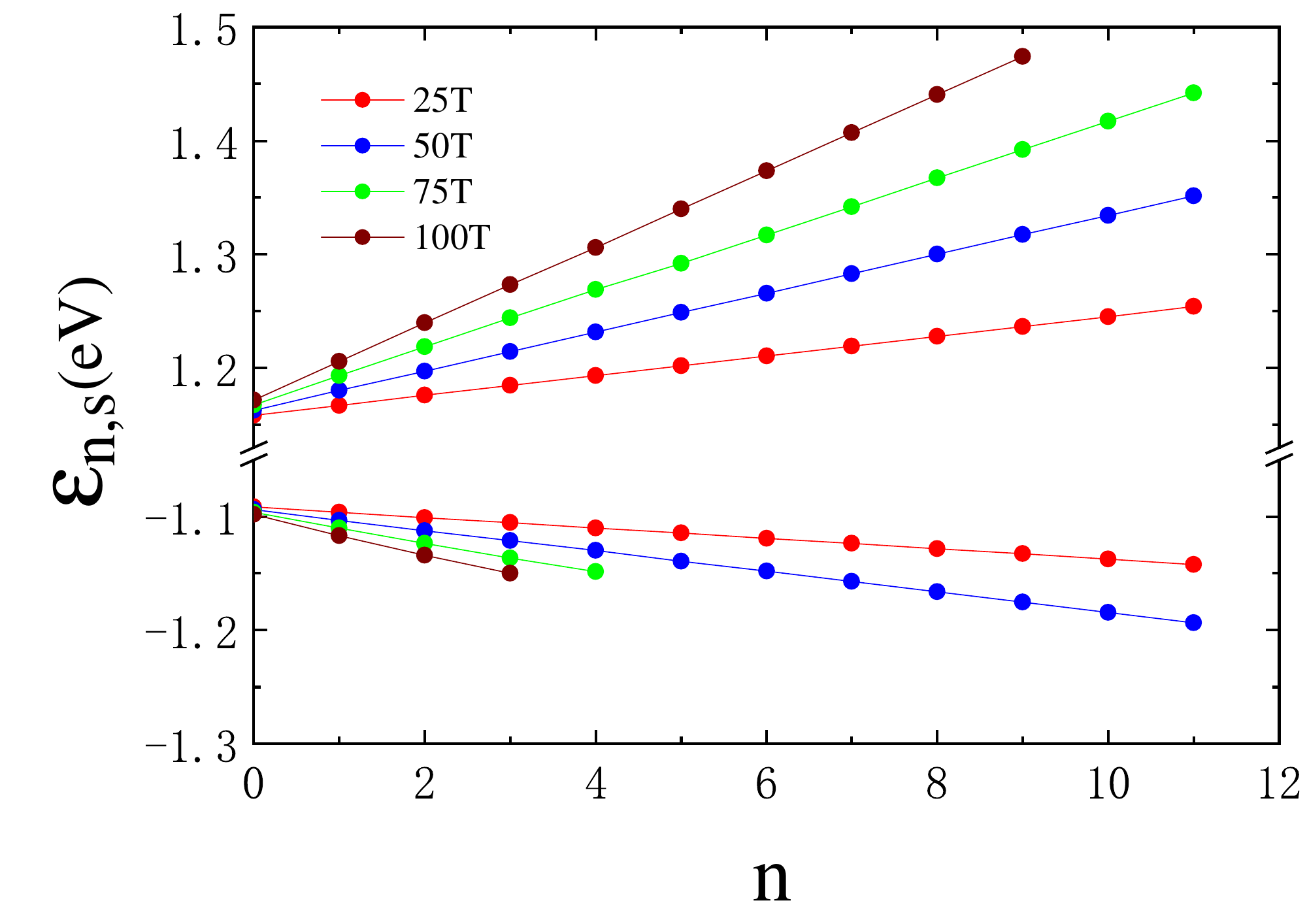}
	\caption{Original (circles) and numerical fitting (line) of Landau levels in single layer SnSe$_2$. For high magnetic field, the landau level has a blue shift and touches $\Gamma$ point. We focus on the states in the low energy part, and thus only present the landau levels in low energy window.}
	\label{fig:label6}
\end{figure}

To go deeper, we investigate the electronic and optical properties of single-layer SnSe$_2$ with external magnetic field, which is still a challenge in DFT calculation. In the case of a  perpendicularly applied magnetic field $B$, the hopping term between two sites $t_{\alpha\beta}$ in Eq.~\ref{Hamiltonian matrix1} is replaced by 
\begin{equation}
t_{\alpha\beta} \rightarrow t_{\alpha\beta}exp[i\frac{2\pi}{\phi_0}e\int_{R_i}^{R_{j}}Adl]
\end{equation}
where $\phi_0$ = hc/e is the flux quantum, and A = (-By, 0, 0) is the vector potential in the Landau gauge, respectively. 
The quantifying magnetic field leads to discrete DOS as the Landau Level (LL) peaks, which are presented in Fig.~\ref{fig:label5}. The geometrical broadening in the LLs is owing to the energy resolution and total number of time steps in the tight-binding propagation method, which is limited by the calculated sample size (number of atoms). Because the electrons and holes are not symmetric in single layer SnSe$_2$, the obtained LL spectrum consists of two sets of equidistant LLs, which can be simply described as 
\begin{equation}
\epsilon_{n,s}^{kp} = E_s +\frac{seB\hbar}{m_e}(n + \frac{1}{2})w_s
\label{landau level}
\end{equation}
Where $s = \pm 1$ denotes the conduction and valence bands, $E_{+/-}$ is the energy at the conduction and valence band edge, $n$ is the Landau index, and $w_s$ = $m_e$/$m_s^\star$ is the relative ratio between free electron and the average effective masses at the band edges. For all investigated magnetic field, the LL spectrum follows well the linear dependence on $n$ in Eq.~\ref{landau level} with fitting values of $E_{+/-}$ = 1.154 (-1.090) eV and $w_{+/-}$ = 2.999 (1.382), respectively, as shown in Fig.~\ref{fig:label6}. On the other hand, the LL spectrum also exhibits linear dependence with $B$, which is similar to a typical 2D electron gas. Unlike the $\sqrt{n}$ in massive Dirac system (sgn(n)$v_f$$\sqrt{2e\hbar B|n|}$), the landau index in Eq.~\ref{landau level} resembles to that of Schr$\ddot{\mathrm{o}}$dinger fermions ($\hbar$$w_c$(n+$\frac{1}{2}$)).

\begin{figure}[ht]	
	\includegraphics[width=8.0 cm]{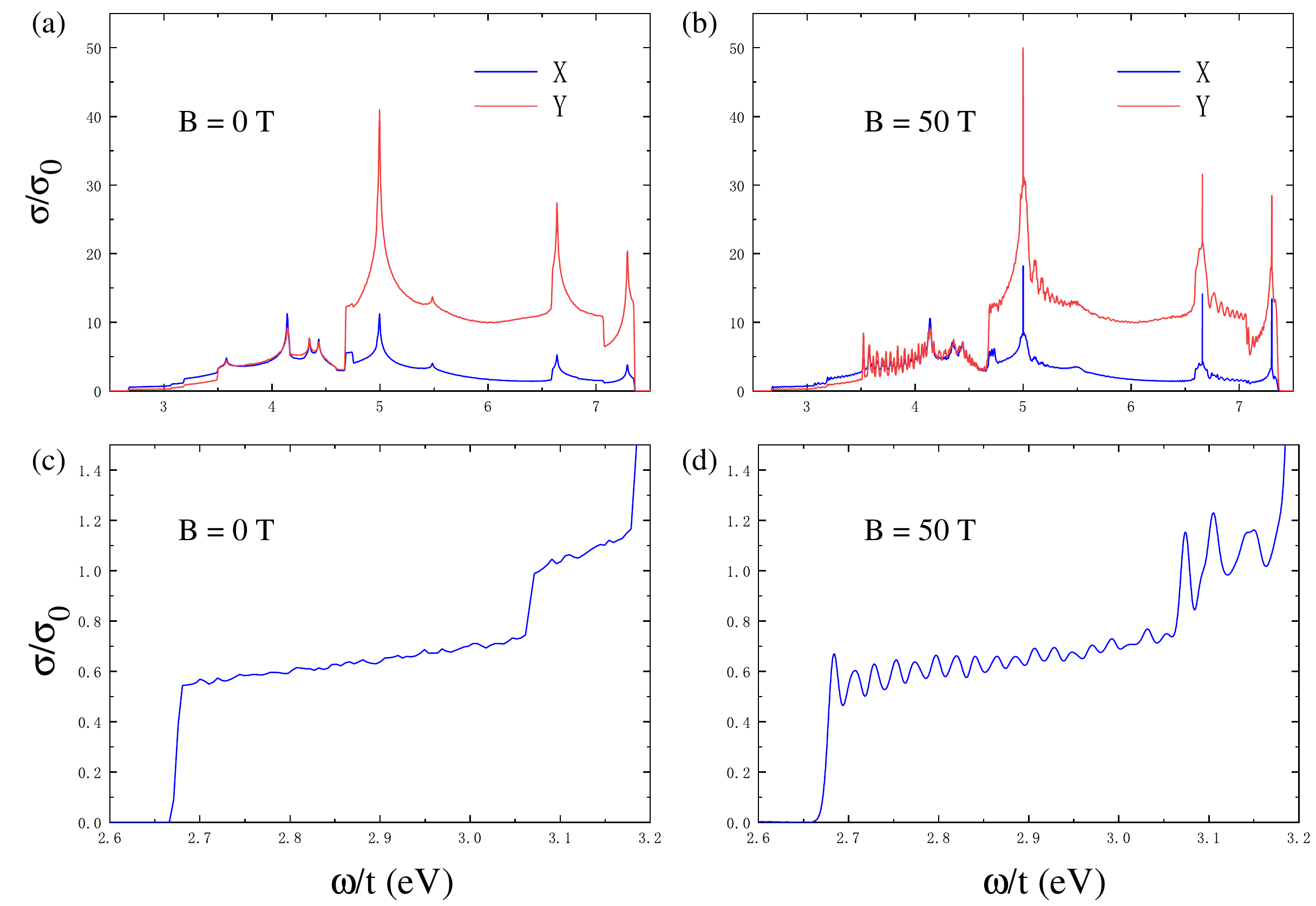}
	\caption{The optical conductivity spectrum of single layer SnSe$_2$ is calculated with B = 0 T (a) and 50 T (b) along X and Y directions, respectively. We zoom in the low energy parts of (a) and (b) along X direction, which are shown at (c) and (d), respectively. $\sigma_0$ = $e^2$/4$\hbar$ is the universal optical conductivity.}
	\label{fig:label7}
\end{figure}

Then, we further study the optical magneto-conductivity $\sigma_{\alpha\beta}$ of single layer SnSe$_2$ using the Kubo formula
\begin{equation}
\begin{aligned}
\mathrm{Re}\sigma_{\alpha\beta}(w) = &\lim_{\epsilon \rightarrow 0^+}\frac{e^{-\beta\hbar{w}}-1}{{\hbar}w\Omega}
 \int_{0}^{\infty}e^{-\epsilon{t}}sin wt \\
& \times 2\mathrm{Im}\left \langle f(H)J_\alpha(t)[1-f(H)]J_\beta \right \rangle dt
\label{optical conduc}
\end{aligned}
\end{equation}
where $\beta=1/k_BT$, $\Omega$ is the sample area, $f(H)=(e^{\beta(H-\mu)}+1)^{-1}$ is the Fermi-Dirac distribution operator, $\mu$ is the chemical potential, and the time-dependent current operator in the $\alpha$ (x or y) direction $J_{\alpha}(t) = e^{iHt}J_\alpha e^{-iHt}$. Fig.~\ref{fig:label7} shows the optical conductivity spectrum of single layer SnSe$_2$ along X and Y directions with and without magnetic field. The sample here is the same as in that the LLs calculation with $\sim$10$^6$ atoms. When no external magnetic field is applied (B = 0 T) in Fig.~\ref{fig:label7} (a), the optical spectrum shows obvious anisotropy, as a result of the anisotropic atomic structure. Specifically, the optical absorption along X direction is strong in the low energy region, but falls down to quarter of that along Y direction in the high energy region. A sharp increase appears at around 2.686 eV, corresponding to the direct optical transition from the highest valence band to the lowest conduction band at M point (2.634 eV). This correspondence confirms the reliability of our proposed TB model again. For the cases of B = 50 T in Figs.~\ref{fig:label7} (b) and (d), the continuous optical conductivity becomes quantized with discrete values, however, which are not very sharp as the cases for graphene and arsene\cite{yuan2010modeling,yu2018tunable}. This is because that the highest valence band at M point is not discrete, clearly shown in the high energy part at -1.50 eV in the black window in Fig.~\ref{fig:label5}(b). Thus, the magneto-optical spectrum can be classified into two categories, one is the direct-gap semiconductors like graphene with absolutely discrete peaks, and the other is indirect-gap semiconductors in this study with no sharp discrete values. Furthermore, the discrete peaks in single layer SnSe$_2$ exhibit blue shifts with an increasing magnetic field. For example, when the magnetic field varies from 50 to 100 T, the first and second peak of the optical conductivity shifts from 2.684 and 2.707 eV to 2.693 and 2.744 eV, respectively.

\section{Summary}
In summary, we have proposed an effective Hamiltonian for single layer SnSe$_2$ that is derived from six orbitals within the self-consistent GW approach. The model shows good performance with respect to the band structure and density of states comparable to those of the GW results, especially, in the low-energy region. Based on the derived TB model, the electronic and optical properties of single layer SnSe$_2$ can be tuned effectively by a perpendicular magnetic field using the time-dependent propagation method. Because of the electron-hole asymmetry, the magnetic field yields two sets of Landau Level spectra. These spectra follow linear dependence on the field strength B and landau index n, confirming the nature of Schr$\ddot{\mathrm{o}}$dinger fermions. Additionally, single layer SnSe$_2$ shows anisotropic optical responses due to its anisotropic atomic configurations. The starting point 
of the optical spectrum (2.686 eV) corresponds well with the quasi-energy gap at M point, validating the reliability of our proposed TB model. In the presence of external magnetic field, the optical conductivity presents some discrete peaks, but very different from those reported in direct-gap semiconductors. The TB model developed in our paper can be used for further exploring the electronic, optical, and transport properties of pristine and disordered SnSe$_2$, especially in the presence of external magnetic fields.

\begin{acknowledgments}
This work is supported by the National Key R$\&$D Program of China (Grant No. 2018FYA0305800). Hongxia Zhong acknowledges the support by China Postdoctoral Science Foundation (Grant No.2018M640723). Numerical calculations presented in this paper have been performed on a supercomputing system in the Supercomputing Center of Wuhan University.
\end{acknowledgments}

\bibliographystyle{apsrev4-1}
\bibliography{references}

\end{document}